\documentclass[aps,twocolumn,prl,showpacs]{revtex4}
\usepackage{graphicx}
\usepackage{dcolumn}
\usepackage{bm}

\newcommand{\n}{$\nu=1$}
\newcommand{\vf}{$V_f$}
\newcommand{\vff}{$V_{2f}$}

\begin{document}

\title{Observation of Chiral Heat Transport in the Quantum Hall Regime}

\author{G. Granger$^1$, J.~P. Eisenstein$^1$, and J.~L. Reno$^2$}

\affiliation{$^1$Condensed Matter Physics, California Institute of Technology, Pasadena, CA 91125
\\
$^2$Sandia National Laboratories, Albuquerque, NM 87185}

\date{\today}

\begin{abstract}
Heat transport in the quantum Hall regime is investigated using micron-scale heaters and thermometers positioned along the edge of a millimeter-scale two dimensional electron system (2DES).  The heaters rely on localized current injection into the 2DES, while the thermometers are based on the thermoelectric effect.  In the $\nu = 1$ integer quantized Hall state, a thermoelectric signal appears at an edge thermometer only when it is ``downstream,'' in the sense of electronic edge transport, from the heater. When the distance between the heater and the thermometer is increased, the thermoelectric signal is reduced, showing that the electrons cool as they propagate along the edge.

\end{abstract}

\pacs{73.43.-f, 73.23.-b, 44.10.+i} \keywords{Quantum Hall Effect, Edge States, Heat Transport}

\maketitle

In the quantized Hall effect (QHE) the interior of the two dimensional electron system (2DES) is incompressible; an energy gap separates the ground state from its charged excitations. Gapless charged excitations do exist, but they are confined to the edges of the 2DES.  These edge excitations are largely responsible for electrical transport through the system. 

Ignoring electron-electron interactions, the gapless edge excitations in integer quantum Hall systems are easy to visualize.  Near the physical edge of the sample the discrete Landau energy levels created by the magnetic field $B$ move up in energy and eventually cross the Fermi level.  Near these intersections the Landau orbitals are unidirectional current-carrying states analogous to classical skipping orbits.  Arbitrarily low energy excitations are possible within each Landau band.  In effect, the 2DES is encircled by a set of chiral, one-dimensional metals, one for each Landau level piercing the Fermi level \cite{halperin}. 

The theory of edge channels in the fractional QHE regime is more complex$~$\cite{KaneFisher1}. Wen$~$\cite{Wen} concluded that the chiral edge states encircling fractional quantum Hall droplets are Luttinger, as opposed to Fermi liquids. For the primitive fractions, $\nu = 1/m$ with $m$ an odd integer, there is a single edge mode propagating in the direction expected for particles of charge $q = -|e|/m$.  For more complex states, such as $\nu = 2/3$, multiple edge modes are expected, some of which propagate {\it upstream} \cite{MacDonald1,Wen,MacDonald2}.  

Understanding the edge of quantum Hall systems is complicated by uncertainty over the sharpness of the edge; i.e.$~$how quickly the electron density falls from its bulk value to zero.  It is widely appreciated that as the edge is softened reconstruction can occur whereby additional pairs of counterpropagating modes appear.  Remarkably, even the \n\ integer quantized Hall state is expected to undergo such an edge reconstruction \cite{Chamon}.

The existence of backward moving modes has yet to be demonstrated experimentally.  Experiments designed to detect backward {\it charged} modes \cite{Ashoori} have so far found no evidence for them.  This motivated us to develop a new means for studying the edge modes of quantum Hall systems, a means not dependent on those modes being charged.  In this paper we report the observation of edge {\it heat} transport in the quantum Hall regime.  Our results demonstrate that at \n\ heat transport is strongly chiral, with heat propagating along the edge of the sample in the same direction as negatively charged excitations.  However, we also find that hot electrons in the \n\ edge channel cool significantly as they propagate.  

\begin{figure}
\includegraphics[width=3in, bb=140 152 459 507]{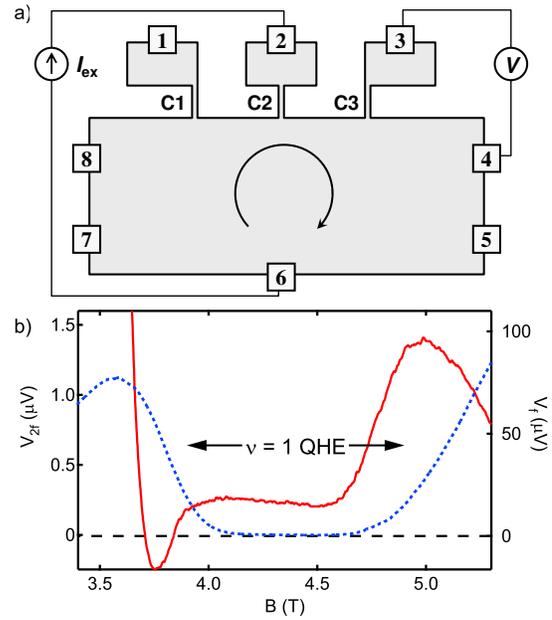}
\caption{\label{} (color online) (a) Schematic diagram, not to scale, of device layout. Numbered squares represent ohmic contacts; C1, C2, and C3 represent constrictions in the 2DES.  (b) \vf\ (dotted) and \vff\ (solid) vs. magnetic field $B$ at $T = 0.1$ K in the \n\ QHE for the measurement circuit indicated. Edge state chirality is clockwise as shown.}
\end{figure}
The 2DES samples employed here are conventional GaAs/AlGaAs heterostructures. The density $N$ and mobility $\mu$ of the 2DES in these samples range from $N = 1.1$ to $1.6 \times 10^{11}$ cm$^{-2}$ and $\mu = 1.6$ to $3 \times 10^6$ cm$^2$/Vs at low temperature.  A schematic illustration, not to scale, of the device geometry is presented in Fig.$~$1a. Diffused NiAuGe ohmic contacts are placed along three of the edges of a large rectangular 2DES.  On the remaining edge (top edge in Fig.$~$1a) three narrow constrictions (C1, C2, and C3) separate the main rectangular 2D region from smaller, but still macroscopic, 2D regions.  Each of these smaller 2D regions has a single ohmic contact.  Devices with two types of constrictions have been studied.  In one case the constrictions are narrow (10 $\mu$m wide, 20 $\mu$m long) channels (NCs) covered by surface gates which control their conductance.  In the other they are quantum point contacts (QPCs) whose conductance is controlled by surface split-gates. Four NC devices and one QPC device, from two different wafers, have all revealed the same qualitative results.  The center-to-center distance between adjacent constrictions, measured along the edge of the main rectangle, is 30 $\mu$m in the NC devices and 20 $\mu$m in the QPC devices.  These constrictions provide a means of locally heating and locally measuring the temperature along the edge of the main 2DES.  The efficacy of this approach was first demonstrated in QPC devices at zero magnetic field by Molenkamp {\it et al.} \cite{Molenkamp}.  

In a typical measurement a low frequency ac excitation current ($I_{ex} \sim 1-50$ nA at $f \sim 5$ Hz) is driven between ohmic contacts 2 and 6 (see Fig.$~$1a) and thus through the center constriction (C2) of the device. If the conductance of this ``heater'' constriction is adjusted (via its associated gates) to be sufficiently small, localized Joule heating of the 2DES in the main rectangle will occur in its vicinity. The resulting temperature rise in the electron gas will extend outward from the constriction a distance determined by various energy relaxation and heat transfer processes.  At low temperatures cooling to the lattice via electron-phonon coupling is weak and this distance can become relatively long.  The existence of chiral edge states in the quantum Hall regime can be expected to significantly impact the extent and directionality of the temperature profile.  

Temperature differences within the 2DES are detected by measuring the voltage difference $V$ between two ohmic contacts, 3 and 4 in this typical example. Contact 3 is attached to the small 2DES region behind constriction C3 (the ``detector'') adjacent to the heater, while contact 4 is attached directly to the main 2DES rectangle.  The voltage difference between these contacts (which, by assumption, are in thermal equilibrium with the lattice) will contain two terms: an ordinary resistive voltage drop and a thermoelectric contribution arising from any temperature drop $\Delta T$ which exists along the constriction.  The existence of this thermoelectric voltage requires only that the thermoelectric power $S$ (Seebeck coefficient) of the detector constriction differ from that of the bulk 2DESs it connects. In order to distinguish the resistive and thermal contributions to $V$, lock-in detection at both the fundamental frequency $f$ and the second harmonic at $2f$ is performed. Since Joule heating is proportional to $I_{ex}^2$, we expect the $2f$ component of $V$ to reflect its thermoelectric component.

We have validated this measurement scheme via experiments performed at zero magnetic field on devices with both NC and QPC constrictions. With C2 used as the heater, clear $2f$ thermoelectric voltages are observed at both C1 and C3.  As observed by Molenkamp {\it et al.} \cite{Molenkamp}, the thermoelectric signal in our QPC device is maximized when the conductance of the QPC detector is on a riser between adjacent quantized conductance plateaus.  That comparable signals are observed with detectors on each side of the heater demonstrates that heat transport at $B=0$ is isotropic as expected and not chiral.

\begin{figure}
\includegraphics[width=3.2in, bb=148 257 443 376]{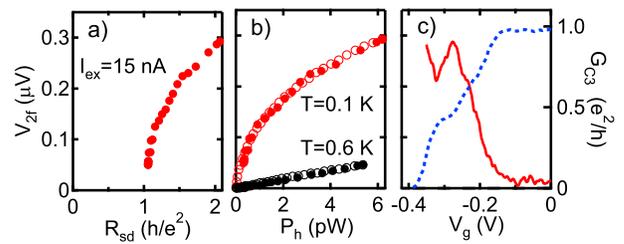}
\caption{\label{}(color online) a) \vff\ vs. net two-terminal resistance $R_{sd}$ of heater circuit at $T=0.1$ K for fixed $I_{ex}=15$ nA. b) \vff\ vs. heater power $P_h$ at $T=0.1$ and 0.6 K.  Closed dots: varying $R_h$ at fixed $I_{ex}$.  Open dots: varying $I_{ex}$ at fixed $R_h$. c) Solid trace: \vff\ vs. gate voltage $V_g$ applied to C3. Dotted trace: $G_{\rm C3}$ vs. $V_g$. All data are at $B = 4.25$ T.}
\end{figure}
Figure$~$1b shows typical results obtained in the vicinity of the bulk \n\ QHE around $B = 4 - 5$ T using a NC device.  A current of $I_{ex} =15$ nA at $f = 5$ Hz is driven between contacts 2 and 6 while both the $f$ and $2f$ components of the voltage difference $V$ between contacts 3 and 4 is recorded.  Constrictions C2 and C3 are adjusted to have conductances of $\sim 0.5$ $e^2/h$ at $B=4.25$ T. Ignoring electron heating, these constrictions would merely add series resistances to the current and voltage pick-up pathways; no effect on the 4-terminal resistivity of the QHE would be expected.  Thus it is not surprising that the resistive component \vf\ of $V$ (dashed trace in Fig$~$1b) shows the deep minimum characteristic of the quantized Hall effect. At the same time, however, a small but non-zero voltage \vff\ is detected at $2f$ (solid trace).  Although the magnetic field dependence of \vff\ is fairly complex on the flanks of the \n\ QHE, we focus here on the center of the state where \vff\ is roughly constant. 

We interpret the $2f$ signal seen within the \n\ QHE state as a thermoelectric voltage arising from a temperature drop along C3 induced by heating at C2.  Support for this interpretation is presented in Fig.$~$2. Figure$~$2a shows how the observed \vff\ signal at $B=4.25$ T depends on the net two-terminal resistance $R_{sd} \equiv V_{sd}/I_{ex}$ (in units of $R_Q=h/e^2$) of the heater circuit.  To obtain these data, the gate voltage controlling C2 is adjusted, and the two-terminal voltage $V_{sd}$ between the source and drain ohmic contacts (2 and 6 in this case) is recorded along with \vff. The excitation current is held constant at $I_{ex} =15$ nA. The figure shows that while \vff\ is non-linear in $R_{sd}$, it appears to vanish as $R_{sd} \rightarrow R_Q$.  This is the expected result.  If the entire heater circuit, including C2, is within the \n\ QHE, then  $R_{sd}=R_Q$.  Heat will be generated, in the amount $P=R_QI_{ex}^2$, but only at hot spots very near the ohmic contacts.  If, as we assume, the ohmic contacts are thermal reservoirs in equilibrium with the crystal lattice, this heat will be absorbed by the contacts.  As the constriction conductance is reduced by gating, $R_{sd}$ starts to exceed $R_Q$.  Additional heating, now in the constriction, begins to occur. As there is no nearby thermal reservoir to absorb this heat, it propagates away from the constriction and is ultimately detected at C3, the detector constriction.  Hence, the data in Fig.$~$2a demonstrate that \vff\ depends not on $R_{sd}$ alone, but rather upon the difference $R_h = R_{sd}-R_Q$, which we may regard as the relevant heater resistance \cite{variability}.

We stress that the data in Fig.$~$2a are obtained at $fixed$ excitation current $I_{ex}=15$ nA.  If \vff\ were a parasitic effect (e.g. harmonic distortion) tied, ultimately, to the resistivity of the 2DES, no dependence on the heater resistance would be expected.  To explore this further, Fig.$~$2b shows the dependence of \vff\ on heater power dissipation, defined as $P_h=R_hI_{ex}^2$, at both $T=0.1$ and 0.6 K.  For both temperatures, the solid dots are obtained by changing the heater resistance at fixed $I_{ex}$, while for the open dots $R_h$ is kept (nearly) constant and $I_{ex}$ is varied (from $I_{ex} = 0.04$ to 15 nA).  To a good approximation, the solid and open dots lie on a single curve.  This shows that \vff\ is a function of heater power $P_h$ rather than another combination of $R_h$ and $I_{ex}$. This is strong evidence in support of our assertion that \vff\ reflects the heating of the 2DES at C2.

The sub-linear power dependence of \vff\ at $T=0.1$ K evident in Fig.$~$2b contrasts with the linear dependence seen at $T=0.6$ K.  This may indicate that at $T = 0.1$ K the electrons in this NC device are being heated well out of equilibrium with the lattice.  Interestingly, we find that in QPC devices comparable \vff\ signals are detectable in the linear regime, even at $T = 0.1$ K \cite{futurepub}. 

Figure$~$2c compares the magnitude of \vff\ at $B=4.25$ T with the conductance $G_{\rm C3}$ of C3, the detector constriction, as functions of the dc voltage $V_g$ applied to the gate across it.  Note that near $V_g=0$, where $G_{\rm C3}=e^2/h$, \vff$\approx 0$.  This is again the expected result since there the 2DES in both C3 and the bulk of the device are within the \n\ QHE state.  The thermopower is therefore uniform along a path connecting contacts 3 and 4 (passing through C3) and thus no thermoelectric voltage can develop \cite{Fletcher}. As $|V_g|$ is increased, $G_{\rm C3}$ falls below $e^2/h$ and \vff\ becomes non-zero. The 2DES in C3 now has (in general) a different thermopower than the bulk 2DES and hence a thermoelectric voltage appears.  We emphasize that the $sign$ of this voltage is consistent with the expected sign (negative) of the thermopower $S$ of C3 and that the electron temperature is higher at the end of the constriction where it meets the large rectangular 2DES than at its other end.  

\begin{figure}
\includegraphics[width=3in, bb=129 107 475 384]{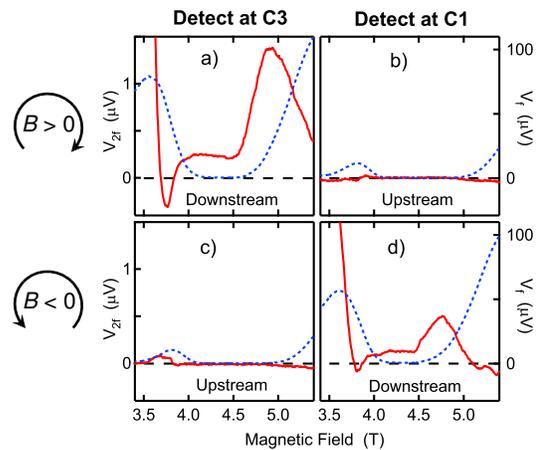}
\caption{\label{}(color online) Chirality of thermal transport at \n\ at $T = 0.1$ K.  a) and d): \vff\ (solid) and \vf\ (dotted) observed downstream from heater constriction, C2. b) and c) \vff\ and \vf\ observed upstream from heater. Edge state chirality and magnetic field directions are indicated.}
\end{figure}
Up to this point the detector, C3, has been downstream (clockwise in Fig.$~$1a), in the sense of electronic edge transport, from the heater, C2. What signals, if any, are observed upstream (i.e.$~$at C1) from the heater?  Figure$~$3 summarizes our findings. Panels (a) and (b) show the resistive, \vf, and thermoelectric, \vff, components of the voltages \cite{voltages} at C3 and C1, respectively, for clockwise edge transport.  Panels (c) and (d) show the same, but for counter-clockwise edge transport (obtained by reversing the magnetic field direction).  The results are unambiguous: while in all four cases the resistive component of the voltage displays the expected QHE minimum, a significant thermoelectric component is only observed downstream from the heater. As expected, therefore, heat transport in the \n\ QHE is $chiral$.  Electrons arriving at an upstream constriction have recently been thermalized at an ohmic contact; those arriving at a downstream constriction have apparently been unable to release the thermal energy they gained in the vicinity of the heater.

How far can hot edge state electrons propagate before they cool appreciably? To investigate this we compared the \vff\ signal observed at C3 when C2 is used as the heater with the C3 signal when C1 is the heater.  (In the latter case C2 is completely closed off by fully depleting the 2DES within it.) In this way we can compare \vff\ signals at a single detector at two different distances from the heater; 20 vs. 40 $\mu$m in the QPC devices and 30 vs. 60 $\mu$m in the NC devices.  In the NC devices only extremely weak thermoelectric voltages could be detected at 60 $\mu$m, suggesting that electrons have almost completely thermalized.  In the QPC devices, a clear signal is observed at 40 $\mu$m, although in the middle of the \n\ QHE it is typically 3 to 5 times smaller than the signal at 20 $\mu$m.  We estimate that the thermal decay length $\lambda$ for hot edge state electrons at \n\ is in the range of $\lambda \sim 20~\mu$m at $T= 0.1$ K.

The 40 $\mu$m \vff\ signal at C3 is reduced if the intermediate constriction, C2, is partially opened.  This behavior is displayed in Fig.$~$4 where we plot the ratio of the 40 $\mu$m \vff\ signal to the 20 $\mu m$ signal (measured separately at C3 with C2 as the heater) as a function of the conductance $G_{\rm C2}$ of C2.  As C2 is opened, a fraction of the hot electrons are diverted away from the edge and are replaced by cold electrons from ohmic contact 2.  The result is a reduced \vff\ signal downstream at C3.  That the signal vanishes at $G_{\rm C2}\approx 0.8~e^2/h$ instead of $e^2/h$ is puzzling. Nonetheless, these data provide strong evidence that in addition to being chiral, heat transport at \n\ is in fact concentrated at the edge of the 2DES.  No analogous quenching of the C3 signal is observed at $B = 0$. 

\begin{figure}
\includegraphics[width=3in, bb=0 0 252 153]{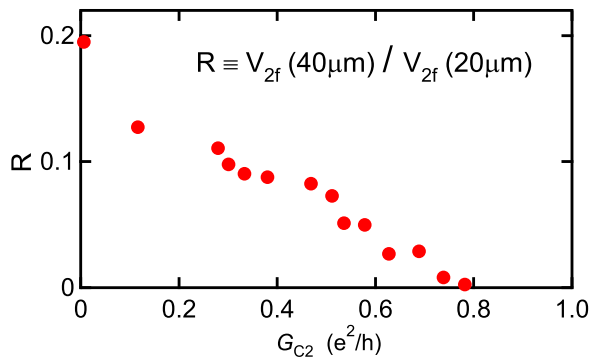}
\caption{\label{}\vff\ signal at $B = 4.25$ T observed with a 40 $\mu$m separation between heater (here C1) and detector C3 vs. the conductance $G_{\rm C2}$ of constriction C2 which lies midway between them. Data normalized by the \vff\ signal observed with a 20 $\mu$m heater-detector separation. $T=0.1$ K.}
\end{figure}
The mechanism responsible for the observed cooling of edge electrons at $\nu = 1$ is so far unknown.  Cooling by acoustic phonon emission is possible, but simple estimates suggest that it is too weak to account for the micron-scale thermal decay length our measurements imply \cite{phonons}. Since the conductivity $\sigma_{xx}$ is vanishingly small in the QHE, naive application of the Wiedemann-Franz law would suggest heat cannot leave the edge and enter the bulk of the 2DES. However, this ignores the possibility of energy transport, mediated by long-range Coulomb interactions, between localized electronic states in the bulk \cite{Bert}. At \n\ such a mechanism might be especially probable given the known existence of low energy neutral collective modes in the spin sector \cite{Cote}.  Another possibility is that there are additional collective modes at the edge itself, due to edge reconstruction \cite{Chamon} or the formation of a compressible strip \cite{Chklovskii}.  For example, a backward moving mode could remove energy from the dominant chiral mode and thus thermalize it. If the backward mode velocity were much less than the chiral mode, little if any heating would be detected upstream.

In conclusion, we have employed local heaters and thermometers to explore heat transport at the edge of the \n\ QHE.  Our results demonstrate that heat transport is strongly chiral but that significant cooling occurs as electrons propagate along the edge.

We thank G. Fiete, M.P.A. Fisher, S.M. Girvin, C.L. Kane, A. Kitaev, A.H. MacDonald, G. Refael, and A. Stern for discussions, and B. Chickering, V. Cvicek, and D. Nichols for technical help. This work was supported via Microsoft Project Q and DOE grant DE-FG03-99ER45766.


\begin{references}
\bibitem{variability} In some cases the heater resistance is better defined as $R_h=R_{sd}-\gamma R_Q$, with $1 \lesssim \gamma \lesssim 1.2$. This seems to imply that a small amount of additional backscattering (and heating) is occuring outside the heater constriction.
\bibitem{futurepub} A detailed description of the power and temperature dependences of the \vff\ signal will be presented elsewhere.
\bibitem{voltages} More precisely, we measure voltage differences between ohmic contacts: for C3 the contacts are 2 and 4; for C1 they are 1 and 8.
\end{references}
\end{document}